# Photoluminescence study of Vanadium in 4H- and 6H- Silicon Carbide

Ibrahim Boulares[1,a], Anne R. Evans[2,b], Jeremy L. Smith[1,c] and Brenda VanMil[1,d*]

[1]DEVCOM Army Research Laboratory, 2800 Powder Mill Road, Adelphi, MD 20783 USA

[2]Department of Electrical Engineering, University of South Florida, Tampa, FL USA

[a]Ibrahim.boulares.civ@army.mil, [b]annee@usf.edu, [c]jeremy.l.smith135.ctr@army.mil and [d]brenda.l.vanmil.civ@army.mil

**Abstract.** Point defects with optically addressable spin states are a key building block for quantum networks, but few solid-state emitters combine a spin-photon interface with emission in the low-loss telecommunications band. Vanadium ($V^{4+}$) defects in silicon carbide (SiC) meet this requirement, emitting zero-phonon lines (ZPLs) throughout the telecom O-band in both the 4H- and 6H-SiC polytypes while retaining an addressable electron and nuclear spin system [1,2]. Although the orbital and spin structure of these defects has been characterized in detail, the polarization selection rules have not been fully mapped experimentally. Here we present polarization-dependent photoluminescence (PL) measurements of vanadium ZPLs in 4H- and 6H-SiC at cryogenic temperature, of two scattering geometries: edge and face-on (emission wavevector perpendicular and parallel to the crystal c-axis, respectively). Emission from two of the four α sites, ($\alpha_2$ and $\alpha_3$) and the 6H-SiC γ sites, are polarized parallel to the c-axis (σ-polarized), whereas emission from α1 and α4 are polarized perpendicular to the c-axis (π-polarized). Additionally, we report preliminary, lower-signal measurements of the β sites.

## Introduction

Solid-state emitters with optically addressable spin states are used in quantum networks, where photons must transmit quantum information over long fiber-optic distances with minimal loss [3,4]. This application is best realized through emission wavelengths within the telecommunications O-, C-, or L-bands (about 1260-1625 nm), where standard silica-fiber attenuation is lowest. Many of the most extensively studied spin-photon defects, such as the nitrogen-vacancy center in diamond, instead emit in the visible range and require frequency conversion for long-distance networking [3].

Silicon carbide (SiC) has emerged as an attractive host for such defects: it is a wide-bandgap, technologically mature semiconductor already used in high-power electronics, is available in large-area wafers, and its two most common polytypes, 4H- and 6H-SiC, host a range of optically active point defects with long spin coherence times [5,6]. Among these, tetravalent vanadium ($V^{4+}$) is notable as the only transition-metal dopant in SiC whose ZPL emission falls entirely within the telecom O-band (1278-1388 nm) in all inequivalent lattice sites of both polytypes [1,2]. Vanadium substituting silicon can occupy two inequivalent sites in 4H-SiC (nominally labeled the α and β centers) and three inequivalent sites in 6H-SiC (labeled α, β, and γ), as established by early PL, magnetic-circular-dichroism, and site-selective ODMR studies [7-9].

The crystal-field level structure of the $V^{4+}$ center was first established by magnetic circular dichroism and site-selective ODMR on the α, β, and γ lines in 6H-SiC [7], and by a detailed Zeeman/crystal-field analysis of the four α transitions ($\alpha_1$, $\alpha_2$, $\alpha_3$, and $\alpha_4$) in 6H-SiC [8]. These works assign the four sub-lines to optical transitions between spin-orbit-split Γ4 and Γ5,6 Kramers doublets derived from the 2E ground and excited orbital states, and establish the electric-dipole selection rules that connect each sublevel pair to a definite linear polarization: the Γ4 to Γ4 transitions carry mixed π (E∥c) and σ (E⊥c) character, while the Γ5,6 to Γ5,6 transitions are purely σ-polarized [7,8]. This π/σ analysis, together with the site labeling above, is what we build on directly in the polarization analysis below.

Detailed ensemble and single-defect studies have established much of the optical and spin structure of the $V^{4+}$ center relevant to quantum technology. Group-theoretical and ab initio calculations and temperature-/time-resolved PL identified the orbital character of the α and β zero-phonon lines and their phonon sidebands in 4H-SiC [1]. Subsequent resonant PL, ODMR, and pulsed spin-resonance measurements resolved the fine orbital structure of all five inequivalent sites across both polytypes, isolated single near-surface $V^{4+}$ emitters via photon antibunching and demonstrated coherent spin control — establishing vanadium in SiC as a viable telecom-band spin-photon interface [2].

What remains comparatively unexplored is the direct, geometry-resolved, mapping of the optical dipole orientation predicted by the site symmetry to the polarization of the emitted photons. In this work, we report polarization-resolved PL measurements of the vanadium ZPLs in 4H- and 6H-SiC using two scattering geometries, edge and face-on, ($\hat{k} \perp \hat{c}$) and ($\hat{k} \parallel \hat{c}$), respectively, where *k* is the excitation/emission wavevector), at cryogenic temperature. The emission from π and σ symmetry transitions may be selectively collected by probing the sample from either the edge or surface, respectively. We use this comparison to test the predicted dipole orientation for the α and γ sites and extend the polarization map to the previously under-characterized β sites.

## Experimental

Measurements were performed on commercial bulk vanadium-doped single-crystal 4H-SiC and 6H-SiC samples, mounted on the cold finger of a closed-cycle cryostat (Montana Instruments) and cooled to a base temperature of approximately 5 K. PL measurements were performed in the back reflection geometry with a 532 nm continuous-wave excitation laser. The PL signal was resolved with an Acton SpectraPro SP-2750 spectrometer and imaged on an InGaAs detector (PyLoN-IR). The polarization of the collected PL was selected via a half waveplate and polarizer installed in the collection path.

In the edge-on geometry, the optical axis is perpendicular to the c-axis, probing the plane in which an in-plane optical dipole is expected to radiate anisotropically as the polarizer angle is rotated. For both geometries, complete PL spectra were recorded as a function of polarizer angle (0-180°) at each resolved ZPL, and the resulting angle-dependent peak areas were used to construct polar plots for each spectral feature.

## Results and Discussion

**Full-spectrum photoluminescence and site identification.** Figure 1 shows low-temperature PL spectra of all five inequivalent sites in 4H-SiC and 6H-SiC. In 4H-SiC, the dominant sharp feature is the α zero-phonon line near 1280 nm, with a weaker β line near 1337 nm. In 6H-SiC, three ZPLs are resolved: α near 1311 nm (the strongest feature), β near 1352 nm, and γ near 1390 nm. These assignments are consistent with the site labeling reported for 6H-SiC:V in [1,2,9] and show that both samples are dominated by well-isolated, spectrally resolved vanadium ZPLs suitable for polarization-resolved study.

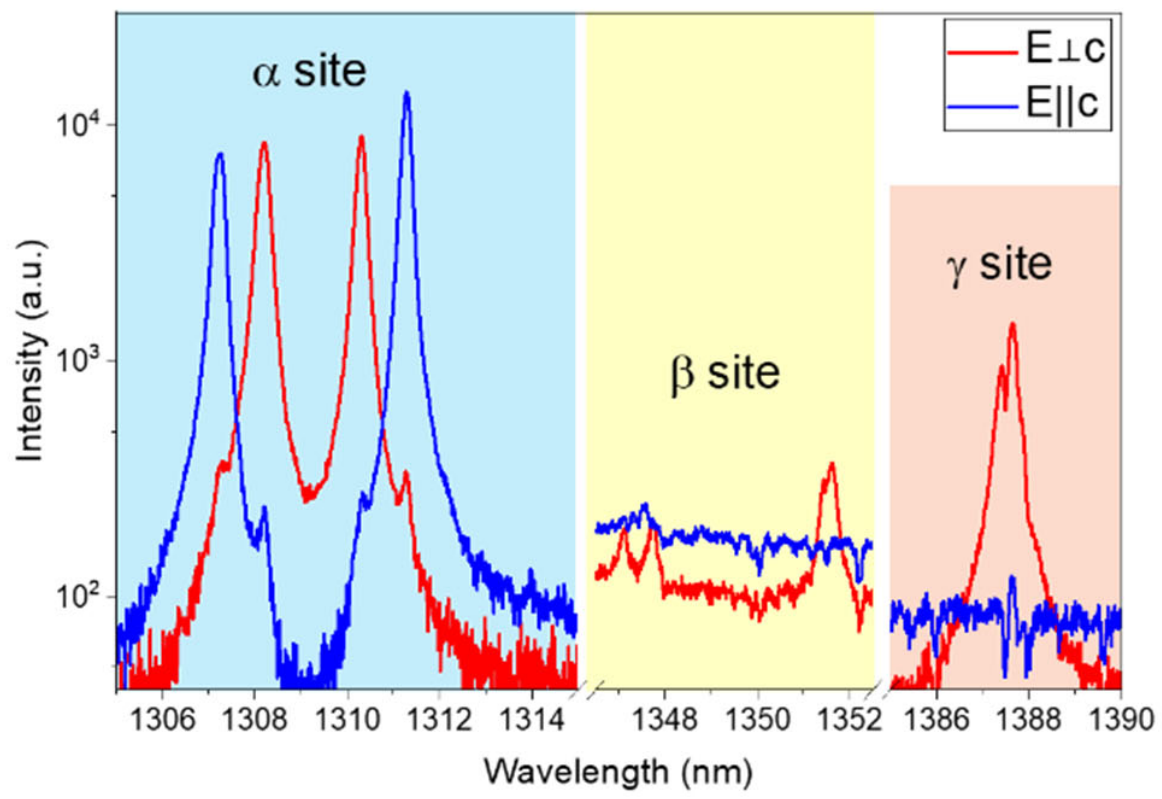

Fig. 1. Low-temperature PL spectra, in the edge-on configuration, of 6H-SiC (left) and 4H-SiC (right), showing the α, β, and (for 6H-SiC) γ zero-phonon lines.

**Polarization dependence of the α zero-phonon line.** Polarization-resolved PL of the α ZPL was recorded at 5 K in both edge-on and face-on geometry for 4H-SiC (1280 nm) and 6H-SiC (1311 nm) and is shown in Figure 2. In the edge-on configuration, all four resolved sub-peaks of the α multiplet (labeled $\alpha_1$-$\alpha_4$) show strong, polarization dependence. The two more intense components ($\alpha_2$, $\alpha_3$) reach maximum intensity at polarization angles rotated by approximately 90° from the two weaker components ($\alpha_1$, $\alpha_4$). This pairing matches the assignment established for the *h*-site α lines in 6H-SiC [7,8], in which $\alpha_2$ and $\alpha_3$ are Γ5,6↔Γ5,6 transitions with pure σ (E⊥c) character while $\alpha_1$ and $\alpha_4$ are Γ4↔Γ4 transitions with mixed π (E‖c) and σ character; since edge-on collection is the only geometry that admits any π component, the different π/σ admixture of the two transition pairs projects onto our rotating polarizer as the 90° offset pattern we observe. In face-on illumination, all four components lose essentially all polarization dependence in both polytypes, appearing as nearly circular polar plots — as expected, since this geometry only collects the σ component, and the σ transitions of a C3v site are not expected to depend on the azimuthal polarizer angle within the plane perpendicular to the c-axis. This edge-on/face-on contrast demonstrates directly that the α transitions follow the π/σ selection rules established by [7,8], consistent with the in-plane-dominated orbital character assigned to this transition [1,2,7,8].

We note one open feature in the face-on data: rather than perfectly circular polar plots, both polytypes show a small but reproducible ellipticity that was reported in [1], as seen in Fig. 2. As a completely in-plane, isotropic dipole viewed exactly along its axis should not exhibit polarization dependence, it is probable that this residual ellipticity is instrumental or strain-related rather than an intrinsic dipole property. Random internal strain has been shown to weakly mix the Γ4 and Γ5,6 sublevels of the 2E states in this system, which would mix a small π-like component into the nominally pure-σ face-on signal [8], and unintended mounting-induced strain has recently been shown to measurably broaden and shift vanadium PLE spectra in similarly mounted, machined SiC samples [10]. A small tilt of the sample c-axis relative to the collection would have a similar effect. Confirming which of these mechanisms dominates will require a controlled tilt or strain study and/or improved sample mounting.

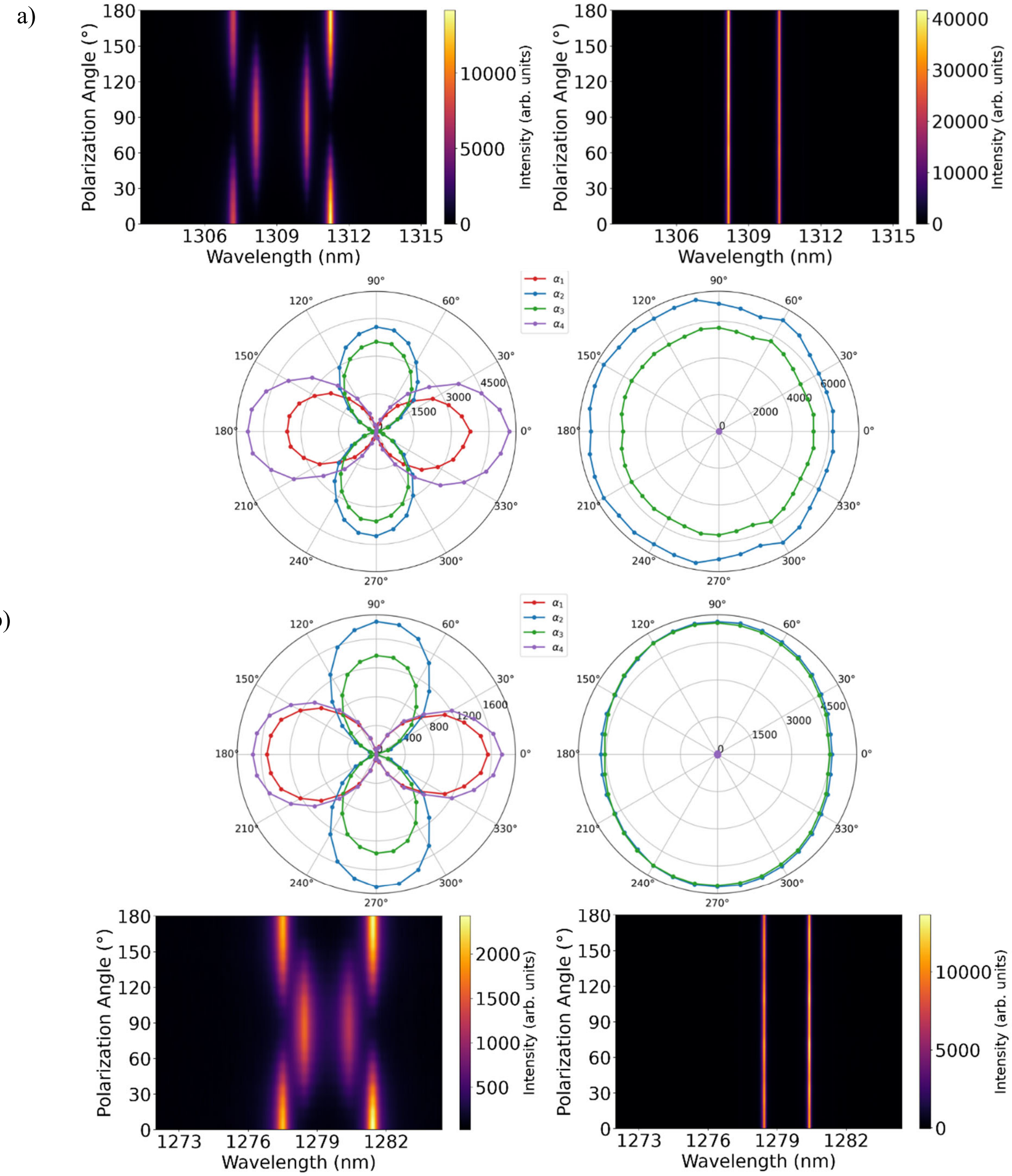

Fig. 2. Polarization-resolved PL of the α zero-phonon line in a) 6H-SiC (1311 nm, top) and b) 4H-SiC (1280 nm, bottom), for edge-on (left) and face-on (right) geometry. Polar plots show fitted peak area of each resolved sub-component (α1-α4) versus polarizer angle and the corresponding polarization-angle heatmap.

**Polarization dependence of the γ zero-phonon line (6H-SiC).** The γ site is unique to 6H-SiC and, at 1390 nm, lies furthest into infrared, making it the most favorable site for low-loss fiber applications among those studied here. Figure 3 shows the same edge-on/face-on comparison for the two resolved γ components ($\gamma_1$, $\gamma_2$). As with the α site, edge-on configuration produces a polarization-dependent pattern consistent with a single Γ pair (rather than two, as for α) contributing measurable intensity. Face-on illumination again shows an approximately polarization-independent response, with the same small ellipticity noted for the α site. Overall, the γ site follows the same qualitative π/σ selection-rule picture established for α [7,8], extending the polarization assignment to a third site.

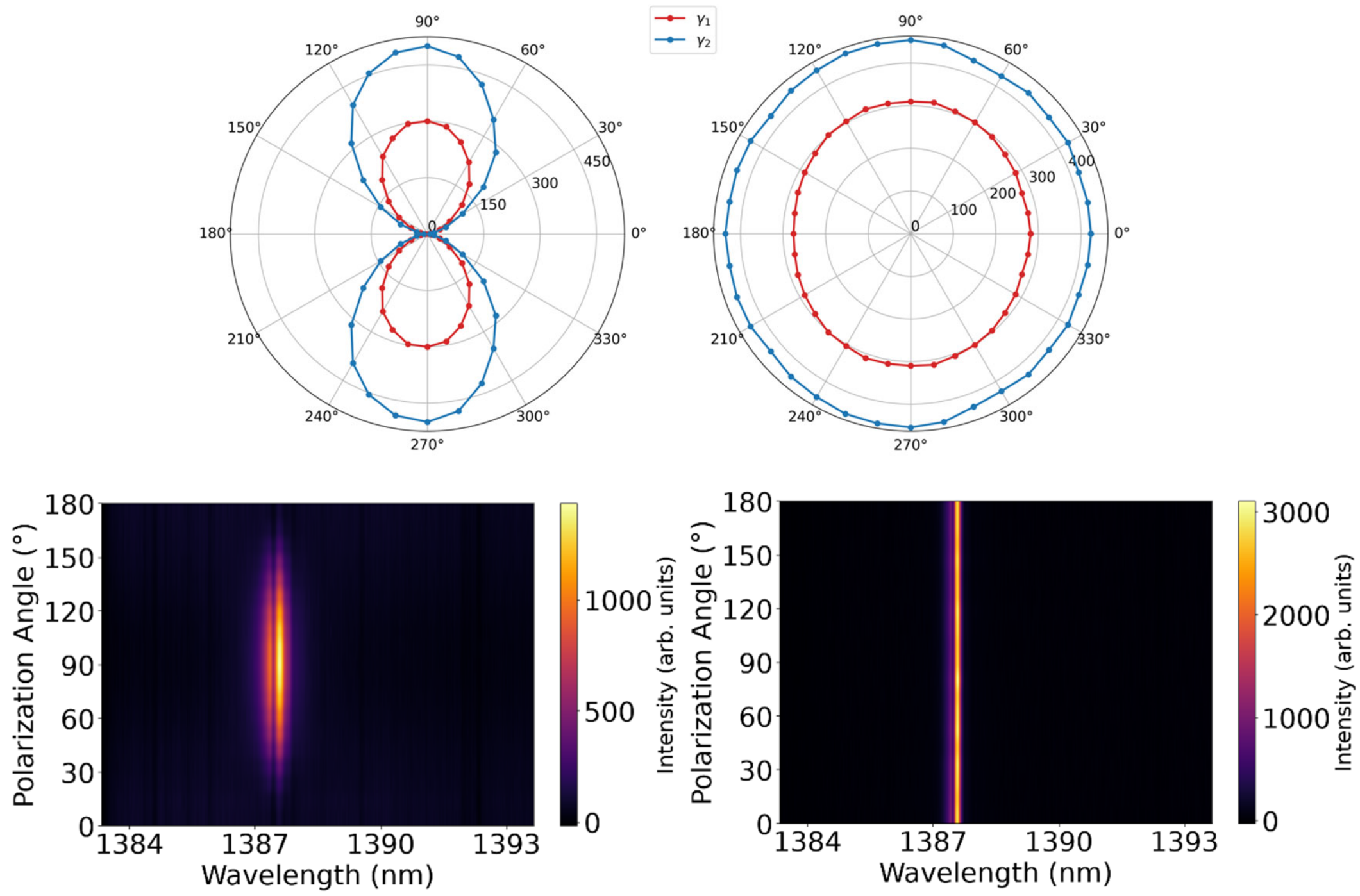

Fig. 3. Polarization-resolved PL and corresponding polarization-angle heatmap of the γ zero-phonon line in 6H-SiC (1390 nm) for edge-on (left) and face-on (right) geometry.

**Preliminary results: the β site.** The β sites in both polytypes were also probed in the same two geometries, but the resulting spectra are considerably weaker and noisier than for α or γ and are reported here as preliminary. Figure 4 shows representative data for the 6H-SiC β line (nominally 1352 nm) in edge-on illumination. Six spectral components are apparent: two better-resolved peaks near 1351-1352 nm, and a cluster near 1346-1348 nm that appears as two broad peaks at low resolution but, on closer inspection (most clearly in the polarization-angle heatmap of Fig. 4), shows shoulders consistent with four unresolved, overlapping components. All six components show clear polarization dependence in edge-on illumination, reaching maximum intensity near 0°/180° and minimum near 90°. However, the signal-to-noise ratio in the present data was not sufficient to obtain reliable fits to the individual peak areas as a function of angle, so a quantitative polar plot analogous to Figs. 2 and 3 could not yet be constructed for this site. The face-on β spectra exhibits a four-component structure without apparent polarization dependence but awaits higher-resolution data before a quantitative comparison can be made. Improving the β-site signal, most likely via longer integration times, higher excitation power at fixed sample temperature, or resonant excitation as used in [2] is a near-term priority for the optical characterization reported here. We note, however, that recent ensemble spin-relaxation measurements find β-site spin-lattice relaxation times below 1 ms, in contrast to the multi-second lifetimes measured for α at sub-Kelvin temperatures [11]. Measurement of edge on polarization showed a very weak signal with a π character, while the face-on resolved two peaks with a σ component.

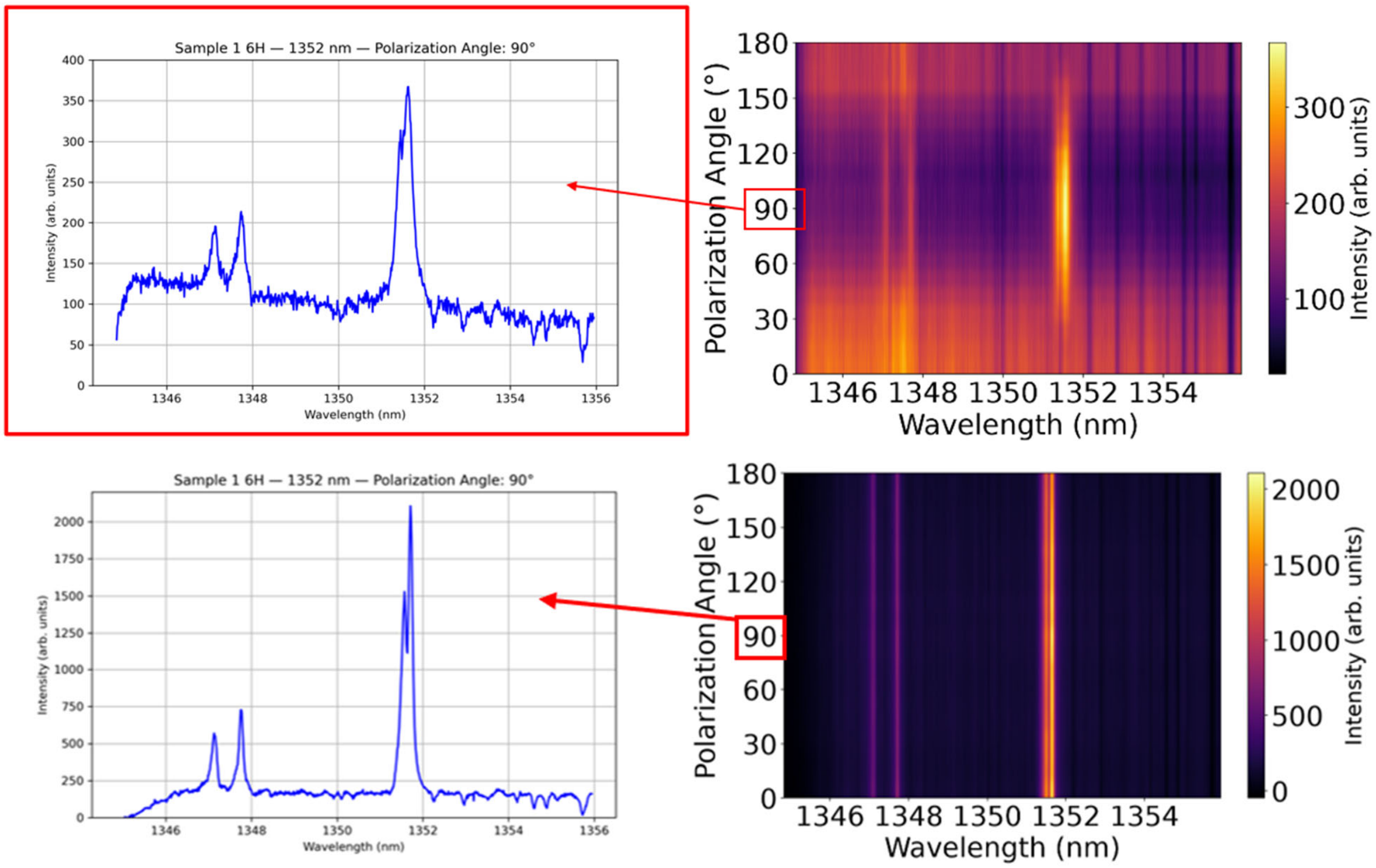

Fig. 4. Preliminary polarization-resolved PL of the β zero-phonon line region in 6H-SiC (Sample 1), edge-on illumination. Top: representative spectra at polarizer angles of 90° (left) and the corresponding polarization-angle heatmap, showing spectrally overlapping peaks (right). Bottom: Face-on spectrum at 90° and corresponding polarization-angle heatmap.

## Summary

We have presented polarization-resolved photoluminescence measurements of vanadium zero-phonon lines in 4H- and 6H-SiC, using two experimental configurations (edge-on and face-on) to directly probe the orientation of the optical transition dipole. For the α and β sites in both polytypes, and for the γ site in 6H-SiC, we observe strongly polarization-dependent emission under edge-on illumination and essentially isotropic emission under face-on illumination, providing direct experimental observation consistent with the site symmetry and with the orbital character established in prior work.